\documentclass[useAMS,usenatbib,RNAAS]{aastex631}
\usepackage{amsmath}

\newcommand   {\sub}[2]  {{#1}_{\mathrm{#2}}}
\begin{document}

\title{Significant improvement in planetary system simulations from statistical averaging}

\correspondingauthor{David M. Hernandez}
\email{dmhernandez@cfa.harvard.edu}

\author{David M. Hernandez}
\affiliation{Harvard--Smithsonian Center for Astrophysics}

\author{Eric Agol}
\affiliation{Astronomy Department, University of Washington}


\author{Matthew J. Holman}
\affiliation{Harvard--Smithsonian Center for Astrophysics} 

\author{Sam Hadden}
\affiliation{Harvard--Smithsonian Center for Astrophysics} 


\begin{abstract}
Symplectic integrators are widely used for the study of planetary dynamics and other $N$-body problems.  In a study of the outer Solar system, we demonstrate that individual symplectic integrations can yield biased errors in the semi-major axes and possibly other orbital elements.  The bias is resolved by studying an ensemble of initial conditions of the outer Solar system.  Such statistical sampling could significantly improve measurement of planetary system properties like their secular frequencies.  We also compared the distributions of action-like variables between high and low accuracy integrations; traditional statistical metrics are unable to distinguish the distribution functions.
\end{abstract}

\section{}

Symplectic maps \citep{hair06} are well suited for studying the astrophysical $N$-body problem; they have been used to study the evolution of our Solar system and exoplanetary systems \cite[e.g.,][]{WH91}.  In this note, we show that an application of a symplectic integrator, regardless of its time step or order, can give biased errors and trajectories.  This bias can be mitigated by considering a statistical sample of trajectories.  

Consider the problem of the Sun plus the four giant planets.  The bodies are treated as point particles obeying Newtonian gravity.  We integrate this system for $10,000$ years using the symplectic Wisdom--Holman map (WH) in canonical Democratic Heliocentric coordinates \citep{WH91,DLL98}.  If $h$ is the timestep and $\epsilon$ is the ratio of the maximum planetary mass to the Solar mass, the error scales as $\mathcal O(\epsilon h^2 )$\footnote{\cite{HD17} used a different error convention for which the scalings in this paper are multiplied by $\epsilon$.}.  As a function of phase space coordinates, the $N$-body Hamiltonian is, 
\begin{equation}
H = \sum_{i}\frac{\vec{P}_i^2}{2m_i} - \frac{Gm_0m_i}{Q_i} + \frac{1}{2m_0}\left(\sum_{i\neq0}\vec{P}_i\right)^2
		- \sum_{0<i<j}\frac{Gm_im_{\!j}}{Q_{i\!j}},
		\label{eq:nbod}
\end{equation}
while the error in the Hamiltonian is \citep{HD17},
\begin{equation}
\begin{aligned}
	\sub{H}{err} &=
	\frac{h^2}{24}\left(   -\frac{1}{m_0}\left(
		\sum_{i\neq0}\frac{Gm_0m_i}{Q_i^3}\vec{Q}_i\right)^2 
		- \sum_{0<i<j}\frac{Gm_im_j}{Q_{i\!j}^5}
	\left[V_{i\!j}^2Q_{i\!j}^2-3(\vec{V}_{i\!j}\cdot\vec{Q}_{i\!j})^2\right]
	+ \sum_{\substack{i,j>0\\i\neq j}}
		\frac{G^2m_0m_im_{\!j}}{Q_{i\!j}^3Q_i^3}\vec{Q}_{i\!j}\cdot\vec{Q}_i
		\right.
		\\
		& \left.
	+\sum_{i\neq0} \left( \frac{G}{Q_i^5} \left[(\vec{P}_{\!\odot}\cdot\!\vec{P}_i)Q_i^2
		-3(\vec{P}_{\!\odot}\cdot\vec{Q}_i)(\vec{P}_i\cdot\vec{Q}_i)\right]
	+ \frac{2 Gm_i}{m_0Q_i^5}
	\left[P_{\!\odot}^2Q_i^2-3(\vec{P}_{\!\odot}\cdot\vec{Q}_i)^2\right] 
	+ \frac{2}{m_i}\left(
		\sum_{j\neq0,i}\frac{Gm_im_j}{Q_{i\!j}^3}\vec{Q}_{i\!j}\right)^2 \right) \right) \\
		&+\mathcal{O}(h^4).
\end{aligned}
\label{eq:hamilt}
\end{equation}
Here, $G$ is the gravitational constant, $\vec{Q}_i$ and $\vec{P}_i$ are the $i$th canonical coordinate vectors, $m_i$ is the $i$th mass, $\vec{V}_i = \vec{P}_i/m_i$, and $\vec{P}_\odot= \sum_{j\neq0}\vec{P}_{\!j}$.  $\vec{V}_{i j} = \vec{V}_i - \vec{V}_j$ and $\vec{Q}_{i j} = \vec{Q}_i - \vec{Q}_j$.  Some terms in eq. \eqref{eq:hamilt} scale as $\epsilon$ and are dominant while others scale as $\epsilon^2$.

We choose $h= 1$ yr and generated output every $10$ years.  The initial conditions are at a time randomly chosen between $0$ to $100$ yrs, $t_{\mathrm{start}}$.  To obtain the initial conditions, we run WH with $17$th order symplectic correctors \citep{WHT96} and $h = 0.005$ yr from time $0$ to $t_{\mathrm{start}}$ (we apply a corrector to the initial conditions and an inverse corrector on output).    Integrating in this way from $0$ to $100$ yrs gives energy error $\delta = \Delta H/H=1.9 \times 10^{-13}$, close to the limit from rounding errors.

\begin{figure}
	\includegraphics[width=200mm]{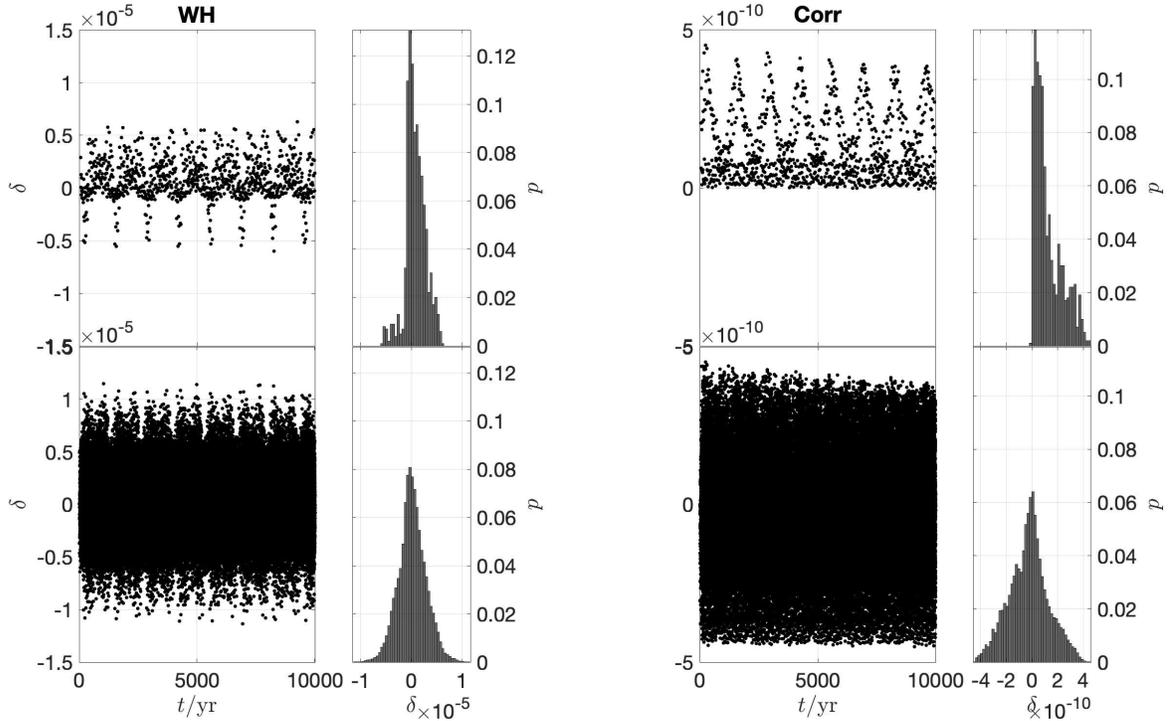}
	\caption{Errors, $\delta$, in integrations of the outer Solar system.  The four left panels show integrations using WH with $h = 1$ yr.  The four right panels show integrations using WH with a $17$th order corrector and $h = 0.01$ yrs.  The top panels show one initial condition.  The bottom panels show the results for $100$ initial conditions; the starting phase for each integration has been chosen randomly.  While the top panels show biased mean energies, the bottom panels mitigate this bias.
		\label{fig:p1}
	}
\end{figure}

The top left panel in Fig. \ref{fig:p1} displays $\delta$ over the $10,000$ yrs.  The error is not  symmetric about $0$, indicating a trajectory with biased errors.  This is confirmed with the adjacent histogram of the errors.  The mean $\delta$, $\overline{\delta}$, is $8.2 \times 10^{-7}$.  This curve is described exactly by the Hamiltonian,
\begin{equation} 
\tilde{H} = H + H_{\mathrm{err}}. 
\label{eq:Htild}
\end{equation}
Note $\tilde{H}$, not $H$, is conserved by the mapping, as can be confirmed numerically \citep{HD17}.

Next, we perform an additional $99$ experiments.  These are identical to the first except that $t_{\mathrm{start}}$ is chosen randomly each time.  The results of the combined $100$ experiments are plotted in the bottom left panel of Fig. \ref{fig:p1}.  The errors are more symmetric about $0$, resulting in $\overline{\delta}=1.5\times10^{-7}$, which is $5.4$ times smaller.  We find evidence that $\overline{\delta}$ is $0$ as the number of initial conditions tends to $\infty$.  For $10,000$ initial conditions, $\overline{\delta }= -7.5\times10^{-9}$, consistent with a standard error scaling as $n^{-1/2}$, and a population $\overline{\delta} = 0$, where $n$ is the number of initial conditions.

The adjacent plot shows a histogram that is approximately normally distributed around $0$ with skewness of $0.042$.  This skewness was not improved by using the 10,000 initial conditions or by allowing $t_{\mathrm{start}}$ to take values between $0$ and $1,000$ years, to capture secular frequency timescales.   

Our experiment demonstrates that the trajectories of a symplectic integrator depend on the initial phase, which is undesired.  We can understand this by noting that the initial phases are found by solving $H$ which, in turn, changes the value of $\tilde{H}$.  The subsequent evolution of $H$ from the mapping depends on the initial $\tilde{H}$.  At leading order in $\epsilon$, the energy error is a weighted sum of the semi-major axes errors, so a biased energy error implies at least one biased semi-major axis error.  Biases in other action-like variables may also exist.

Our averaging procedure improved the distribution of errors.  {Such averaging procedures could have a significant impact, for instance, in improving calculation of secular frequencies in the Solar System \citep{Laskar2011}.}  

Substantial effort has been invested into higher accuracy symplectic integrators \citep{Y90,WHT96,LR01,Blanesetal2013,Reinetal2019b}.  We investigate also their biases, using WH with a $17$th order symplectic corrector.  The error scales as $\mathcal O(\epsilon h^{18})$ + $\mathcal O(\epsilon^2 h^2)$.  The integration time remains the same; output is generated every $10$ yrs, and the timestep is $h = 0.1$ yrs.  The right panels of Fig. \ref{fig:p1} show the results for one and $100$ initial conditions, respectively.  Again, histograms of the errors are also shown.  The error is dominated by the $\mathcal O(\epsilon^2 h^2)$ term.  Compared to the low accuracy experiments, the order in the error has changed by $\epsilon (0.1)^2 = 10^{-5}$, explaining the change in scale.  This is a factor $\epsilon$ smaller than if we had just reduced the time step.  Our goal is to reduce the error, which we have done by using a smaller step and symplectic correctors, but other ways of reducing error exist.

For one initial condition, the error is again not symmetric about $0$, and $\overline{\delta} = 1.2 \times 10^{-10}$.  When the $100$ initial conditions are studied, $\left|\overline{\delta}\right|$ decreases by a factor $5.0$; $\overline{\delta}$ is now $-2.5 \times 10^{-11}$.  The errors are approximately normally distributed around $0$ with skewness of $0.022$.  An ensemble of initial conditions again improved bias for higher accuracy methods.  

Using standard statistical tools to distinguish action-like variables between the low and high accuracy integrations is not helpful.  We computed the probability distribution functions (PDFs) in eccentricity and semi-major axis of Jupiter for the $100$ low accuracy and $100$ high accuracy integrations.  Performing a 2-sample Kolmogorov-Smirnov test finds the distributions to be statistically consistent at the $5\%$ significance level.  Small differences in the PDFs due to the accuracy of the integrator are not statistically significant.  Statistical agreement between disparate integrators in Solar System dynamics has also been shown by \cite{Hetal2020}.

High accuracy symplectic integrators have a number of disadvantages compared to conventional integrators.  The high accuracy symplectic integrators,
\renewcommand\labelenumi{(\theenumi)}
\begin{enumerate}
\item can at best only match the error behavior of conventional methods obeying Brouwer's Law \citep{HMR07,HernandezHolman2021,BartramWittig2021},
\item do not readily accomodate adaptive stepping, which can make computation more efficient, and
\item unlike conventional integrators, typically assume a conservative Hamiltonian system is being solved.
\end{enumerate}

Our results lead us to conclude that we can achieve a high degree of precision and accuracy in measurement of some Solar system quantities through ensembles of low accuracy integrations.




\bibliographystyle{aasjournal}
\bibliography{rnaas}


\end{document}